# A System for Compressive Sensing Signal Reconstruction

Irena Orovic, Andjela Draganić, Nedjeljko Lekić and Srdjan Stanković

University of Montenegro, Faculty of Electrical Engineering, 81000 Podgorica, Montenegro, email: irenao@ac.me

[1]*Abstract-* **An architecture for hardware realization of a system for sparse signal reconstruction is presented. The threshold based reconstruction method is considered, which is further modified in this paper to reduce the system complexity in order to provide easier hardware realization. Instead of using the partial random Fourier transform matrix, the minimization problem is reformulated using only the triangular R matrix from the QR decomposition. The triangular R matrix can be efficiently implemented in hardware without calculating the orthogonal Q matrix. A flexible and scalable realization of matrix R is proposed, such that the size of R changes with the number of available samples and sparsity level.**

**Index Terms— Compressive sensing, hardware realization, QR decomposition, threshold, scalable architecture**

## I. INTRODUCTION

Compressive sensing (CS) [1], [2] is the area dealing with reduced set of signal samples. It has been shown that the signals, having concise representation in a certain transform domain, can be fully reconstructed using just few randomly selected available samples. The process of signal reconstruction plays a crucial role in CS. To that end, various powerful and complex reconstruction methods are defined, with the aim to achieve an accurate signal reconstruction and to cover a wide range of applications (radar systems, multimedia, communications, biomedicine). The implementation of the CS algorithms would reduce the required number of sensors and the energy consumption. Commonly used algorithms could be classified through the following groups: basis pursuit (based on $l_1$ minimization), greedy algorithms such as orthogonal matching pursuit and iterative thresholding algorithms. Most of these algorithms require a significant number of iterations for high reconstruction accuracy, which increases the execution time. The threshold based single iteration algorithm was considered earlier as appropriate for hardware implementation due to its simplicity along with high reconstruction efficiency [3], [4]. This algorithm can be used to recover missing signal parts incurred either as a consequence of CS strategy or as a result of L-estimate filtering. In this paper, we propose further simplification of the original algorithm, in order to improve the architecture given in [4]. Particularly, a complex and numerically demanding part used for threshold calculation is modified and reduced to the constant (under certain conditions). The main part of the hardware consists of a system that selects appropriate elements of the full Fourier transform matrix in order to create partial random Fourier transform (PRFT) matrix. Namely, the PRFT matrix is obtained by selecting rows and columns that correspond to the positions of spectral components and available signal samples, respectively. The spectral components positions are obtained by applying the threshold, derived to separate signal components from the spectral noise caused by missing samples. Another challenge for the hardware implementation of the considered reconstruction algorithm is the least square problem solution using matrix inversion operation that is commonly solved using QR decomposition. In this paper we provide the solution for the least square optimization using only triangular R matrix, thus avoiding computation of matrix Q. Due to the properties of R matrix, this solution significantly simplifies the implementation of the system. Furthermore, as the size of matrix R should change with the number of available samples and signal sparsity level, a flexible and scalable solution is provided allowing variable matrix sizes. Beside the system architecture, all blocks within the scheme are discussed.

The paper is structured as follows. In Section II, a theoretical background about the threshold based single iteration algorithm is presented. The system implementation is proposed and discussed in Section III. The calculation complexity is analysed in Section IV, while the conclusion is given in Section V.

[1] This work is supported by the Montenegrin Ministry of Science, project grant: “New ICT Compressive sensing based trends applied to: multimedia, biomedicine and communications”.

## II. THEORETICAL BACKGROUND

A simple solution that provides efficient reconstruction of sparse signals with missing samples is proposed in [3]. The Fourier domain is used as a domain of signal sparsity, but the approach can be extended to other transform domains. It is based on the threshold that selects the spectral components of the signal in just one iteration. As a result of missing samples, the spectral noise appears in the Fourier domain. Based on the analysis of noise in the spectral domain, the threshold separating signal and noise components is derived. Consider a signal $x(n)$ having $K$ components in the discrete Fourier transform domain, where $K<<N$ ($N$ is the total signal length). According to the CS theory, this signal can be reconstructed from much fewer samples than required by the Nyquist Shannon theorem. Thereby, a random undersampling is required. The missing samples will cause noise in the spectral domain with an approximate variance [3]:

$$\sigma^2 = M\frac{N-M}{N-1}\sum_{a=1}^{M} x^2(n_a)/M = \mu\sum_{a=1}^{M} y^2(a), \qquad (1)$$

where $n_a$ denotes the positions of available samples, $y(a)=x(n_a)$ are the available samples called measurements, while the constant $\mu=(N-M)/(N-1)$. The noise variance depends on the number of missing samples $M$, i.e., the number of available samples $N-M$. This variance is crucial for the threshold calculation:

$$T = \frac{1}{N}\sqrt{-\sigma^2 \ln(1-P^{1/(N-K)})} \approx \frac{1}{N}\sqrt{-\sigma^2 \ln(1-P^{1/N})}\ , \quad (2)$$

which provides that all $(N-K)$ noise components are below $T$, with the given probability $P$. The Fourier transform calculated using only available samples is called the initial Fourier transform $\mathbf{X}$ and it is obtained as:

$$X(k) = \sum_{a=1}^{M} y(a)e^{-j2\pi n_a k/N}, k = 1,...,N\ . \qquad (3)$$

Therefore, the positions of coefficients in (3) that are above the threshold $T$ define the frequency support of signal components, obtained as follows:

$$\mathbf{k} = \arg\{|\mathbf{X}| > T\}\ . \qquad (4)$$

After the signal support is determined, the exact amplitudes of the Fourier coefficients need to be calculated. For that purpose a minimization problem has to be solved. Namely, from the full size Fourier transform matrix $\mathbf{A}_{N\times N}$, we form CS matrix $\mathbf{A}_{CS}$, whose rows correspond to the positions $n_a$ of available samples, while the columns correspond to the signal components support $\mathbf{k}$. As a solution of the optimization problem we obtain:

$$\hat{\mathbf{X}} = (\mathbf{A}_{CS}^{*}\mathbf{A}_{CS})^{-1}(\mathbf{A}_{CS}^{*}\mathbf{y}), \qquad (5)$$

where $\mathbf{y}$ is a vector of available measurements: $\mathbf{y}=\mathbf{x}(n_a)$, $a=1,\ldots,M$, while $\mathbf{A}_{CS}^{*}$ is Hermitian transpose of $\mathbf{A}_{CS}$. The solution of (5) provides the exact amplitudes of $K$ signal components in the Fourier transform domain.

## III. Block scheme for Hardware architecture

A proposed architecture for improved hardware realization of the signal reconstruction algorithm discussed in the previous section is shown in Fig. 1. *Block 1* determines positions of the spectral components above the threshold. Observe that the FFT block is used for the initial Fourier transform calculation. Then, the matrix $\mathbf{A}_{CS}$ is created within the *Block 2*.

*Block 3* performs the calculation of triangular $\mathbf{R}$ matrix (part of the QR decomposition) corresponding to the matrix $\mathbf{A}_{CS}$. The optimization problem is solved within this block as well.

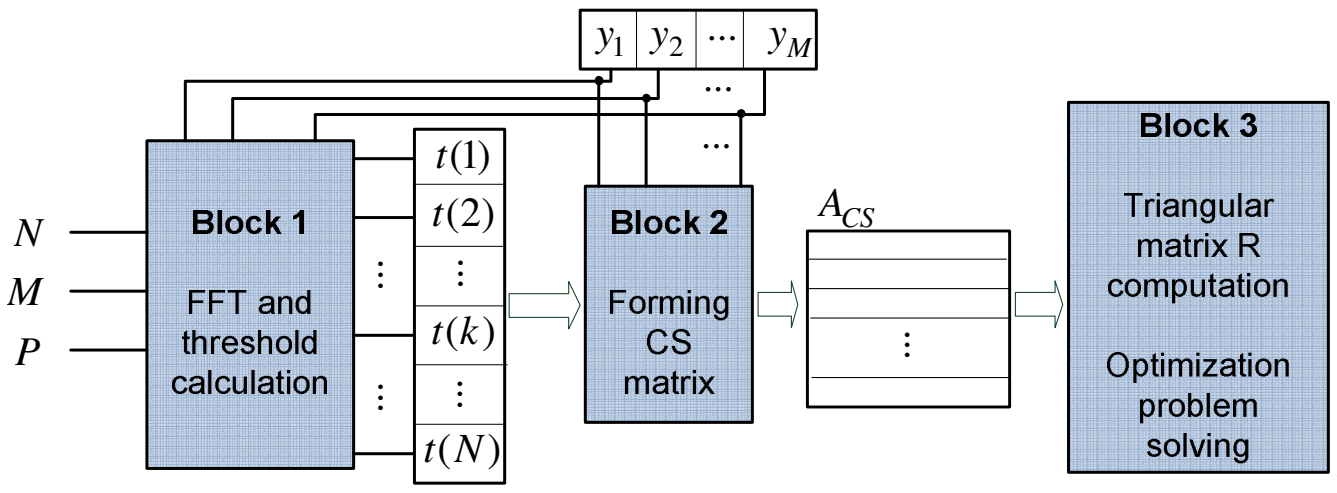

Fig. 1: Block scheme for the single iteration reconstruction algorithm hardware

*A.* ***Block 1*** (Fig. 2) consists of the part for initial Fourier transform calculation by using FFT routine based devices [6]. The threshold is calculated in the parallel branch with the FFT and its absolute values. Note that the FFT coefficients are complex and therefore, the ABS circuits includes the calculation of the square root of the squared real and imaginary parts. The absolute values of the initial Fourier transform and threshold $T$ are fed to the input of the comparators block (COMP). At the output of the COMP block, we obtain the vector $\mathbf{t}$ that consists of logical values 0 and 1. Values 1 are obtained at the position where the absolute values of FFT coefficients are above the threshold T, otherwise, the logical zeros are obtained. The most challenging part within this block is an appropriate realization of the threshold. Namely, according to (2), we would need devices for logarithm, signal power and square root calculations, which require significant hardware complexity. Instead, we provide an optimal system design that would be simple and suitable for real-time applications. Thus, we introduce certain approximations obtained from the real-case analysis. In many applications, it is possible to set a fixed value for the probability such as $P$=0.99. Now, we can assume that for different $N$:

$$\ln(1-P^{1/N}) \approx const. = C\ , \qquad (6)$$

and consequently the calculation of the threshold $T$ in (2) can be simplified as follows:

$$T = \frac{1}{N}\sqrt{-\sigma^2 C}\ , \qquad (7)$$

where $\sigma^2$ is calculate using (1).This further means that the most demanding operation for threshold calculation is the square root (Fig. 2). In the literature, there are various realizations for the square root calculation: Babylonian method, Newton-Raphson method, Taylor-series expansion algorithm, restoring and non-restoring algorithm. Here we will consider the square root calculation using the non-restoring algorithm, described in the sequel [9]. The square root of 2*n* bit number, requires *n* cycles through datapath while the number of gate counts is around 34.38*n*.

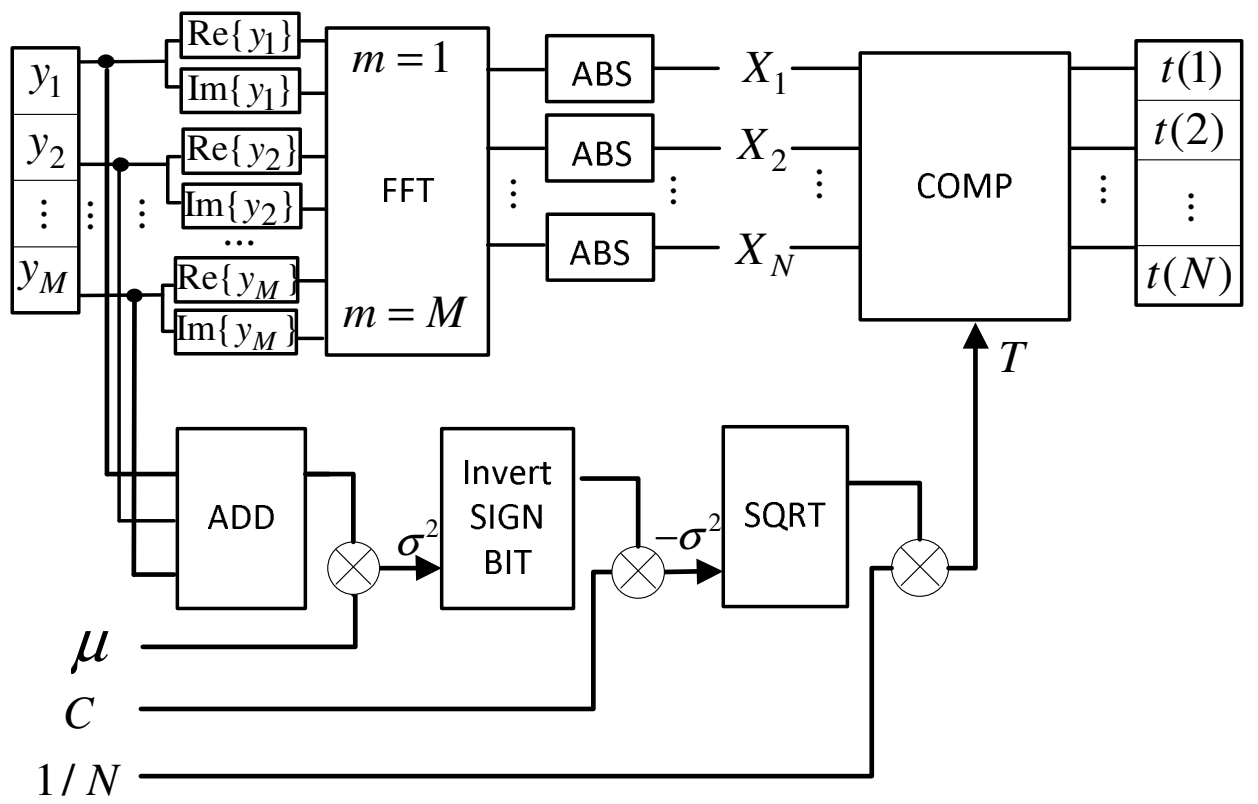

Fig. 2: *Block 1* consists of the part for FFT calculation, threshold computation and comparator block which determine signal support in the frequency domain

The algorithm is based on the recursive relation: $s_i = 2s_{i-1} \pm (2D_{i-1} \mp 2^{-i})$, where $s_i$ is the *i*-th partial remainder, while $D_i$ is the square root up to the *i*-th digit. Square root bit $d_i$ is determined based on the value of partial remainder. If $s_i>0$ then $d_{i+1}$ equals 1 and $(2D_{i-1}+2^{-i})$ is subtracted from the $2s_i$. Otherwise, $d_{i+1}$ equals -1 and $(2D_{i-1}-2^{-i})$ is added to the $2s_i$. Consider the square root of *B* and *B*>1 in our case. The algorithm can be summarized through the following steps:

1. $s_0=B-1$, $D_0=1$;
2. $s_i = 2s_{i-1} \pm (2D_{i-1} \mp 2^{-i})$;
3. $s_i > 0,\ d_{-(i+1)} = 1;\ \ s_i < 0,\ d_{-(i+1)} = -1$.

The root digits are from the set {-1,1} (e.g. *D*=1.1-11-111-1), but can be written using the set {0,1} (e.g. *D*=1.0101101).
For implementation, square root block needs adder/subtracter, combinatorial logic, registers and shift registers.

***B. Block 2*** The second part of the system architecture in Fig. 1 refers to the PRFT matrix. Using the elements of the full Fourier matrix we need to select columns that correspond to the signal components and rows corresponding to the positions of available samples. In other words, CS matrix $\mathbf{A}_{CS}$ is formed by choosing the elements at the intersections of the corresponding rows and columns. If the Fourier transform matrix is stored in memory in row-major order, then the elements of $\mathbf{A}_{CS}$ matrix are selected using the memory addresses in the form:

$$address = (n_a - 1)N + b(j),\ \ n_a \in \{n_1, ..., n_M\},\ \ j \in \{1, ..., K\}\ ,(8)$$

where $\boldsymbol{b}$ keeps the positions of Fourier transform components, that are above the threshold, i.e. positions of elements $\boldsymbol{t}(k)=1$ $k=1,\dots,N$ at the output of COMP circuit (Fig. 1).

***C. Block 3*** represents the system architecture for the optimization problem solving. The main challenge is realization of the inverse matrix $\mathbf{A}_{CS}$. Usually the inversion is done using QR decomposition into matrices $\mathbf{R}_{CS}$ and $\mathbf{Q}_{CS}$ [10],[10]. After the QR decomposition is applied, the optimization problem is recast to the form:

$$\mathbf{X} = \left(\mathbf{A}_{CS}^*\mathbf{A}_{CS}\right)^{-1}(\mathbf{A}_{CS}^*\mathbf{y}) = \left((\mathbf{Q}_{CS}\mathbf{R}_{CS})^*(\mathbf{Q}_{CS}\mathbf{R}_{CS})\right)^{-1}(\mathbf{A}_{CS}^*\mathbf{y}) \qquad (9)$$

where the matrix $\mathbf{Q}_{CS}$ is orthogonal i.e. ${\mathbf{Q}_{CS}}^*\mathbf{Q}_{CS} = \mathbf{I}$ and the matrix $\mathbf{R}_{CS}$ is triangular. Namely, using the orthogonality property of matrix $\mathbf{Q}_{CS}$, we can simplify (9) as follows:

$$\mathbf{X} = \left(\mathbf{R}_{CS}^{-1}({\mathbf{R}_{CS}}^{-1})^*\right)\cdot\left(\mathbf{A}_{CS}^*\mathbf{y}\right). \qquad (10)$$

Therefore, we avoid calculation of the matrix $\mathbf{Q}_{CS}$ and the inversion of $\mathbf{A}_{CS}$ is reduced to the inversion of $\mathbf{R}_{CS}$, which is far less demanding (computation complexity for $\mathbf{R}_{n\times n}$ matrix inversion is $n^2$).

Let us now consider the realization of matrix $\mathbf{R}_{CS}$ and its inverse. There are several methods for decomposing certain matrix into triangular and orthogonal one: Gram-Schmidt decomposition; Householder transformation and Givens rotations (GR). Here, we consider the GR approach, since it provides the best trade-off between accuracy and computational complexity [10],[12], while the realization can be parallelized. The $\mathbf{R}_{CS}$ matrix computation, based on the input matrix $\mathbf{A}_{CS}$, is shown in Fig. 3a. Firstly, the matrix is rearranged and fed to the QR decomposition cells (marked with *r*). Each column of the rearranged matrix has one element delay in compare to the previous one. It is represented as zero input (Fig. 3a). As a consequence, for the matrix of size $M\times K$, the $M+(K-1)$ steps are required for $\mathbf{R}_{CS}$ computation. The diagonal (circle) cells produce coefficients $c_i$ and $s_i$, i.e., the coefficients of the GR. For the element of the input matrix $\mathbf{A}_{CS}$ denoted by $r_{in}$, the rotation coefficients are calculated as:

$$c_i = \frac{r_{i,j}}{\sqrt{r_{i,j}^2 + r_{in}^2}},\ \ s_i = \frac{r_{in}}{\sqrt{r_{i,j}^2 + r_{in}^2}}, \qquad (11)$$

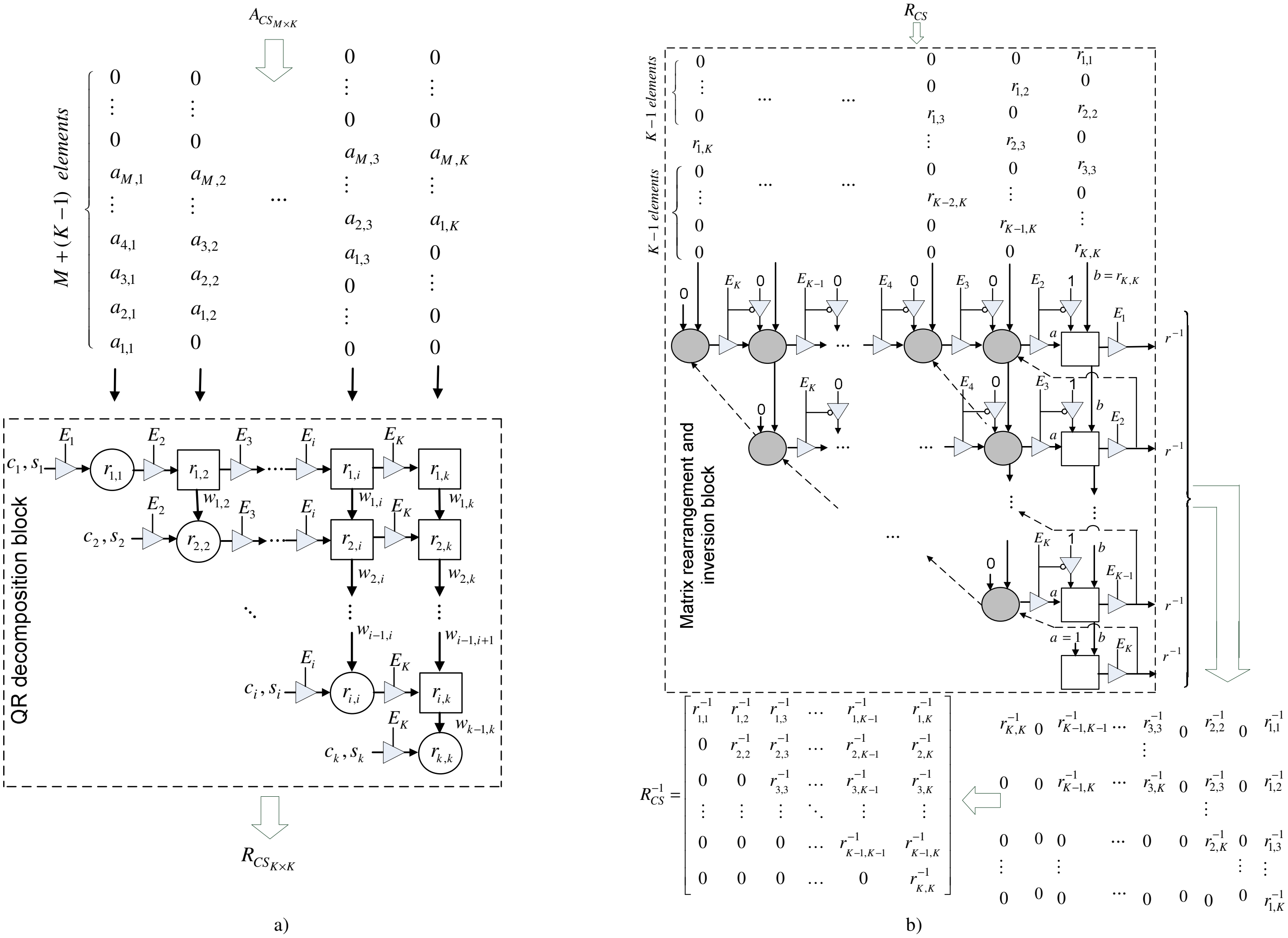

Figure 3: a) QR decomposition block; b) Matrix inversion block

where $c_i$ and $s_i$ are rotation coefficients calculated in the current cell, $r_{i,j}$ is the current value in the cell, which is updated after $c_i$ and $s_i$ calculation:

$$r_{i,j} = \sqrt{r_{i,j}^2 + r_{in}^2}\,. \tag{12}$$

Inner (rectangular) cells update their values according to:

$$r_{i,j} = c_{in} r_{i,j} + s_{in} r_{in}\,, \tag{13}$$

where $c_{in}$ and $s_{in}$ denote coefficients at the input of rectangular cells. Note that $c_{in}$ and $s_{in}$ coefficients at the input of the rectangular cells do not correspond to $c_i$ and $s_i$ coefficients calculated at the circle cell of the corresponding row. The rectangular cells receive delayed rotation coefficients, where delay increase with the distance between rectangular and circle cell. At the outputs of the rectangular cells, the $w_{out}$ is produced according to the relation:

$$w_{out} = -\mathrm{s}_{in}\, r_{i,j} + c_{in} r_{in}\,. \tag{14}$$

As a result, in the final step each cell contains one value that corresponds to the value of the triangular matrix $\mathbf{R}_{CS}$, while the values below the matrix diagonal are zeros.

Since we generally need to work with matrices of different sizes, we need to provide a flexible architecture (Fig. 3a) that allows having arbitrary number of available samples $M$ and signal components $K$. Initially, systolic array has $N$ circular cells and $(N^2-N)/2$ rectangular cells, where $N>M$ and $N>K$ holds. This is the maximal size of the systolic array. As different signals express different sparsity $K$ and different number of measurements $M$ can be used from the signal, the systolic array has to be of adjustable size. In that sense, the control signals $E_1$:$E_N$ are introduced. They operate on matrix columns and turn off the column cells, whose control signals are set to "0". Therefore, for $\mathbf{A}_{CS_{M\times K}}$, ($M$>$K$) systolic array from Fig. 3a will have first $K$ circular cells and ($K^2$-$K$)/2 rectangular cells. The rest of the circle and rectangular cells should be turned off (by setting $E_i$=0, for $i$=$K$+1,…,$N$).

After matrix $\mathbf{R}_{CS}$ is obtained, it is rearranged and zero padded, then fed to the block for triangular matrix inversion proposed in Fig. 3b. Having matrix of $K\times K$ dimension, the inversion is done in 2$K$-1 steps. The input of the rectangular (circle) cell can be either value "1" ("0"), or element of the triangular matrix, or the output of the neighboring cell. Similarly as in the previous case, systolic array initially has $N$ rectangular cells and ($N^2$-$N$)/2 circular cells. The enable signals $E_i$=0 turn off the cells whose index $i$ is $i$>$K$ (Fig. 3b). Rectangular cells output the elements of the inverted matrix. The outputs are obtained as: $r^{-1}$=-$a$ /$r_{i,i}$ (Fig. 3b):

$$r_{i,i}^{-1} = -1 / r_{i,i},\ r_{i,j}^{-1} = -u_{i,j} / r_{i,i}, \tag{15}$$

where $r_{i,i}$ is a diagonal element, $a$ can be -1 or $u_{i,j}$, $j$=$i$+1,…, $K$; and $u_{i,j}$ is calculated as:

$$u_{i,j} = \sum\nolimits_{k=1}^{j-1} r_{i,k} r_{k,j}^{-1}. \tag{16}$$

and $r_{i,k}$ is a value inside the cell, while $r_{k,j}^{-1}$ is an inverted matrix element coming to the cell from the diagonal (Fig. 3b).

## IV. Calculation complexity

Let us discuss first the calculation complexity of the GR method. It performs a set of plane rotations of the original matrix, in order to decompose matrix into two parts – orthogonal matrix **Q** and upper triangular matrix **R**. For the rectangular matrix $\mathbf{A}_{m\times n}$, where $m$>$n$, GR method requires $3n^2(m-n/3)$ floating point operations (flops) for calculation of matrix **R**, while **Q** matrix calculation requires $4n(m^2-n^2/3)$ flops. Since in the proposed solution we do not calculate **Q**, it means that a number of $4n(m^2-n^2/3)$ flops for its calculation are avoided.

Let us now discuss complexity of the proposed solution for the least square problem. The multiplication of complex matrices of sizes $p\times n$ and $n\times m$ requires 4$pmn$-2$pm$ additions and 4$pmn$ multiplications 0. Having matrices of size $\mathbf{R}_{CS_{K\times K}}^{-1}$, $\mathbf{A}_{CS_{K\times M}}^{*}$ and vector $\mathbf{y}_{M\times 1}$ in (10), solving least square problem according to this relation requires $2K(K^2+K+2M-2)$ additions and $4K(K^2+M+K)$ multiplications. For comparison, approach defined by (5) requires $2K(2MK-K+2M-1)$ additions and $4MK(K+1)$ multiplications, while complexity is additionally increased for $K^2$ for inversion of the product matrix ($\mathbf{A}^*_{CS}\mathbf{A}_{CS}$). For example, let us observe the real-world case: $K$=15 signal components and $M$=250 available samples. Using the proposed system, the total number of additions is 22140 and total number of multiplications is 29400. However, by using (5), the total number of additions is 239520, and total number of multiplications is 240000, which is much more operations compared with proposed form (10).

The number of clock cycles to perform reconstruction is:

a) 3$M$+61 for THRESHOLD block,
b) $O(\log M)$ for comparator circuit,
c) 126$K$-8 for matrix **R** calculation block,
d) 34$K$-4 for matrix inversion, and
e) $O(KM+K^2)$ for optimization problem solving.

## V. Conclusion

A system realization for threshold based CS algorithm performing the reconstruction of randomly undersampled signals is considered. The modification of the reconstruction algorithm is proposed aiming to adapt the algorithm for hardware implementation. The calculation of the logarithm and power function is avoided and approximated by a constant term. Also, the minimization problem is refined using only an upper triangular matrix R calculated form the CS transform matrix. The proposed system is suitable for different CS matrix sizes, different sparsity levels and different number of available samples. Another advantage of the proposed system is scalability, achieved by introducing the control signals for the elements of the systolic arrays.